\documentclass[aip, reprint, twocolumn,jcp,english]{revtex4-1}
\usepackage[T1]{fontenc}
\usepackage[latin9]{inputenc}
\usepackage{amsmath}
\usepackage{amssymb}
\usepackage{comment}
\usepackage[citecolor=blue]{hyperref}
\usepackage[framemethod=tikz]{mdframed}
\makeatletter
\usepackage{tikz}
\usetikzlibrary{calc}
\usetikzlibrary{arrows}
\usetikzlibrary{shapes.geometric}
\tikzset{
    vertex/.style={circle,draw,minimum size=1.5em},
    edge/.style={->,> = latex'}
}
\usetikzlibrary{shapes.misc, positioning}
\usetikzlibrary{decorations.markings}
\usepackage{appendix}

\global\long\def\ket#1{\left|#1\right\rangle }%

\global\long\def\bra#1{\left\langle #1\right|}%
 
\global\long\def\braket#1#2{\left.\left\langle #1\right|#2\right\rangle }%
\global\long\def\orthrep#1{\widetilde{#1}{\!\;\!}}%
\global\long\def\dotorthrep#1{\dot{\widetilde{#1}}{\!\;\!}}%
\global\long\def\tder#1{\dot{#1}{\!\;\!}}%
\makeatother
\usepackage{babel}

\begin{document}

\title{Time Evolution of ML-MCTDH Wavefunctions I: Gauge Conditions,
Basis Functions, and Singularities}
\begin{abstract}
We derive a family of equations of motion (EOMs) for evolving multi-layer multiconfiguration time-dependent Hartree (ML-MCTDH) wavefunctions that, unlike the standard ML-MCTDH EOMs, never require the evaluation of the inverse of singular matrices.   All members of this family of EOMs make use of alternative static gauge conditions than that used for standard ML-MCTDH.  These alternative conditions result in an expansion of the wavefunction in terms of a set of potentially arbitrary orthonormal functions, rather than in terms of a set of non-orthonormal and potentially linearly dependent functions, as is the case for standard ML-MCTDH.  
We show that the EOMs used in the projector splitting integrator (PSI) and the invariant EOMs approaches are two special cases of this family obtained from different choices for the dynamic gauge condition, with the invariant EOMs making use of a choice that introduces potentially unbounded operators into the EOMs. As a consequence, all arguments for the existence of parallelizable integration schemes for the invariant EOMs can also be applied to the PSI EOMs.
\end{abstract}
\author{Lachlan P Lindoy}
\email{ll3427@columbia.edu}
\author{Benedikt Kloss}
\author{David R Reichman}
\affiliation{Department of Chemistry, Columbia University, 3000 Broadway, New York, New York 10027, USA}
\maketitle

\section{Introduction}
The multiconfiguration time-dependent Hartree (MCTDH)\citep{MEYER199073,MANTHE1992,BECK20001}
and its multi-layer extension (ML-MCTDH)\citep{WANG2003,MANTHE2008,VENDRELL2011,WANG2015}
have found considerable success in obtaining systematically
convergable dynamics for large quantum mechanical systems. Due to
their general applicability, these approaches have been successfully employed to 
simulate the dynamics of small isolated systems \citep{MANTHE1992,MANTHE199212,MANTHE19937,HAMMERICH1994,RAAB1999,VENDRELL20071,VENDRELL20072,VENDRELL20073,VENDRELL20074,WORTH2008,MENG2013} and considerably larger condensed-phase systems.\citep{CRAIG2007,WANG2007,KONDOV2007,CRAIG2011,BORRELLI2012,RABANI2015,SCHULZE2016,SHIBL2017,MENDIVE2018}
As these approaches can, in principle, be converged to give the exact
quantum dynamics, the results obtained using these techniques are also often used to benchmark other approaches.\citep{Wang2010,Miller2018,Richardson2020} 
However,
despite these successes there are regimes in which such approaches
fail to capture the correct dynamics.\citep{HINZ2016,MEYER2018148,MEYER2018149}
One important cause for such difficulties is that the underlying 
equations-of-motion (EOMs) can become singular or nearly singular, necessitating 
the use of regularization techniques.\citep{BECK20001} While regularization does
not lead to significant numerical difficulties in many applications, in some cases
regularization can dramatically influence the accuracy and efficiency
of the approach, and these issues are not always easily detectable.\citep{MEYER2018148,MEYER2018149,Weike2021}
Problems with regularization 
have been found to become more significant as the number of physical
degrees of freedom in the system increases.\citep{WANG2021,Weike2021}
For the very large scale calculations that are now accessible
these problems need to be addressed.

The singularity in the ML-MCTDH EOMs arises due to a fundamental
issue associated with any approach that makes use of the ML-MCTDH
ansatz, or more generally any tensor network representation which uses
a time-dependent variational principle (TDVP) to obtain EOMs. Whenever
the ML-MCTDH wavefunction is rank deficient, meaning that the exact wavefunction
can be expressed using fewer expansion coefficients than are present
in  the ML-MCTDH wavefunction employed, the TDVP does not uniquely define
the evolution of all coefficients in the ML-MCTDH wavefunction. In
order to truly resolve this issue and obtain unique, non-singular EOMs,
it is necessary to go beyond the linear variations employed within
the TDVP. However, while approaches
have been developed that use such techniques,\citep{MANTHE2015,MANTHE2018279,Yang2020} 
they can involve considerably
more effort than approaches based on linear variations.

Even if the ML-MCTDH wavefunction is not exactly rank deficient,
it can often be nearly rank deficient during the early stages of the simulation.
In the standard ML-MCTDH approach, this can lead to ill-conditioning
of the EOMs, and it becomes necessary to regularize the EOMs. In challenging
regimes, convergence with respect to the regularization parameter can
be difficult to achieve, and this has motivated approaches that attempt
to resolve this problem.  One such approach that has found success in improving the stability
of the ML-MCTDH EOMs makes use of an improved regularization scheme which leads to considerably more stable dynamics, and has allowed for
dynamics to be obtained for models that are otherwise inaccessible
via standard ML-MCTDH.\citep{MEYER2018148,MEYER2018149,WANG2021}

An alternative means to alleviate
the issues associated with rank deficiency and near rank deficiency
involves the use of a different representation of the ML-MCTDH
wavefunction, one in which the singularities that can be observed
in the ML-MCTDH EOMs do not appear. Two approaches of this class are the invariant
EOMs \citep{Weike2021} and the projector splitting integrator (PSI) method.\citep{LUBICH2015,LUBICH2016,KLOSS2017,BONFANTI2018252,CERUTI2021}
The PSI has found success in the simulation of the unitary dynamics
of wavefunctions represented using Matrix Product States,\citep{Schroeder2016,LUBICH2016MPS,KLOSS2018,PAECKEL2019}
and the MCTDH \citep{KLOSS2017} and ML-MCTDH \citep{Schroder2019,KLOSS2020,Bauernfeind2020,CERUTI2021}
ans\"{a}tze. The EOMs that are treated within this approach are non-singular
regardless of whether the ML-MCTDH wavefunction is rank-deficient.
However, this approach does not fundamentally resolve the issue associated
with rank-deficiency of the ML-MCTDH wavefunction. In particular,
rank-deficiency here manifests as non-uniqueness in the alternative representations
that are used.\citep{MEYER2018148,KLOSS2017} As a result,
there have been questions raised within the ML-MCTDH community as
to how well such an approach can work,\citep{MANTHE2015,MEYER2018148}
and there has been no general and systematic study of the relationship between
these different representations or an understanding of the conditions for which individual approaches are optimal.

In this series of papers, we aim to address these concerns, and then demonstrate
the tangible numerical benefits that arise from these considerations.
In this work we will present a unified framework for deriving the
ML-MCTDH EOMs, the PSI EOMs, and the invariant EOMs, as a means of
clarifying the relationships between these approaches. In particular,
we will discuss how the choice of gauge condition used within the
standard ML-MCTDH approach results in an expansion of the wavefunction
in terms of non-orthogonal and potentially linearly dependent sets
of functions for all non-root nodes, giving rise to the singularities in the final EOMs. We will show
that the PSI EOMs and invariant EOMs both make use of the same alternative
representation in which the wavefunction is expanded in terms of a
set of orthonormal but potentially arbitrarily selected functions. This is
the source of the issues relating to the non-uniqueness of these approaches.
These representations introduce an additional gauge freedom
with respect to the standard ML-MCTDH EOMs from which we will derive the EOMs for generic
dynamic gauge conditions, referred to as the Singularity
Free EOMs. From this we will show that the PSI EOMs and invariant
EOMs simply arise from different choices for this dynamic gauge condition.
As such, the parallelizable integration scheme employed for the invariant
EOMs\citep{Weike2021} can readily be extended to treat the PSI EOMs,
and thus there is nothing inherent to the latter approach that restricts it to serial
updates. However, in contrast to the PSI EOMs which are always non-singular,
the choice for the dynamic gauge condition that is implicitly made
in the original derivation of the invariant EOMs (and which we will
make explicit) leads to singularities and therefore the need for
regularization. 

This paper is structured as follows. In Sec.\,\ref{sec:The-ML-MCTDH-Wavefunction}
we will review the ML-MCTDH representation of the wavefunction and
present the notation that will be used throughout. In Sec.\,\ref{sec:The-ML-MCTDH-Equations},
we will present a derivation of the standard ML-MCTDH EOMs that considerably
simplifies the process of arriving at singularity-free EOMs and provides a convenient intermediate point for obtaining singularity-free EOMs. In
Sec.\,\ref{sec:Singularity-Free-Equations}, we will present
the alternative representation of the wavefunction that is used by
the PSI and invariant EOMs, and by using this representation derive
a set of EOMs that are free of singularities and describe the evolution
of the full ML-MCTDH wavefunction. Some additional discussion of these EOMs is provided in the Supplementary Information. 
In Sec.\,\ref{sec:The-Invariant-Equations},
we will discuss the invariant EOMs approach, and how it relates to
the singularity-free EOMs.  We will present our conclusions
in Sec.\,\ref{sec:Conclusion}.

In the companion paper we will aim to address concerns about
the numerical performance of the approaches we advocate. We will discuss the implementation
of the PSI for the ML-MCTDH wavefunction, with a focus on how the
steps required relate to those required to implement standard ML-MCTDH.
A series of applications of the PSI to models
that represent significant challenges to standard ML-MCTDH, even when
applying recently proposed improved regularization schemes, are then presented. These
models will include a simple two-mode model that highlights the issue
of rank deficient wavefunctions, a series of spin-boson models that
have previously been considered with standard ML-MCTDH\citep{MEYER2018148,MEYER2018149,WANG2021} using
both standard and improved regularization schemes with baths containing
up to $10^6$ bath modes, and a series of multi-spin-boson models
that require considerably larger (in terms of size of coefficient
tensors) ML-MCTDH wavefunctions to obtain converged results. 

\section{The ML-MCTDH Wavefunction\label{sec:The-ML-MCTDH-Wavefunction}}

\begin{figure}
\centering
\includegraphics{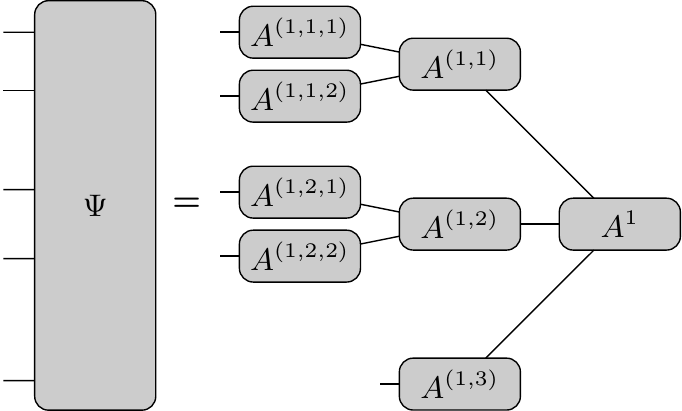}
\caption{\label{fig:mctdh_wavefunction}A diagram representing a ML-MCTDH approximation
to a five-dimensional wavefunction. Each node of the tree corresponds
to a coefficient tensor, and the lines connecting nodes represent
the contraction over indices common to the two nodes.}
\end{figure}

In the ML-MCTDH approach, the high-dimensional time-dependent
wavefunction $\left|\Psi(t)\right\rangle $ is expanded in terms of
a basis constructed as a direct product of $d^{1}$ sets of ``single-particle
functions'' (SPFs), $\left|\phi_{i}^{(1,k)}(t)\right\rangle $, as\citep{WANG2003,MANTHE2008,VENDRELL2011}
\begin{equation}
\left|\Psi(t)\right\rangle =\sum_{i_{1}}^{n_{1}^{1}}\dots\sum_{i_{d^{1}}}^{n_{d^{1}}^{1}}A_{i_{1}\dots i_{d^{1}}}^{1}(t)\bigotimes_{k=1}^{d^{1}}\left|\phi_{i_{k}}^{(1,k)}(t)\right\rangle .\label{eq:wavefunction_expansion_spf}
\end{equation}
Each of these SPFs can then be expanded in terms of a direct
product basis of $d^{(1,k)}$ sets of SPF for a further reduced subset
of degrees of freedom. Applying this process recursively, each of the $N^{z_{l}}$
SPFs associated with a node can be expressed in terms of a direct
product of SPFs as\citep{WANG2003,MANTHE2008,VENDRELL2011}
\begin{equation}
\left|\phi_{i}^{z_{l}}(t)\right\rangle =\!\sum_{i_{1}}^{n_{1}^{z_{l}}}\!\dots\!\sum_{i_{d^{z_{l}}}}^{n_{d^{z_{l}}}^{z_{l}}}\!A_{i_{1}\dots i_{d^{z_{l}}}i}^{z_{l}}(t)\bigotimes_{k=1}^{d^{z_{l}}}\left|\phi_{i_{k}}^{(z_{l},k)}(t)\right\rangle,
\end{equation}
where the label $z_{l}=(b_{1}\dots b_{l})$, contains the $l$
indices $b_{i}$ that denote the path required to reach this SPF starting
from Eq.\,\ref{eq:wavefunction_expansion_spf}, and $d^{z_{l}}$ is
the number of sets of SPFs forming the direct product basis. This process
is repeated until at the bottom level the SPFs are expanded
in terms of a primitive set of basis functions for the physical degrees
of freedom of the system
\begin{equation}
\left|\phi_{i_{k}}^{(z_{l},k)}(t)\right\rangle =\Big|\chi_{\alpha}\Big\rangle.
\end{equation}
From these definitions, the wavefunction can be expressed in terms of a 
series of contractions between the coefficient tensors, $A_{i_{1}\dots i_{d^{z}}i}^{z_{l}}(t)$. This process
leads to a wavefunction that can be represented by a tree structure, as shown in Fig.\,\ref{fig:mctdh_wavefunction}, and is referred to as a tree tensor network.  
The coefficient tensors are represented
as nodes of the tree, and contractions over indices common to two tensors
are represented by lines linking the nodes. The ML-MCTDH wavefunction
can, in principle, exactly represent an arbitrary wavefunction, and
the accuracy of this expansion depends on the number of SPFs used
at each node.

\begin{figure}
\centering
\includegraphics{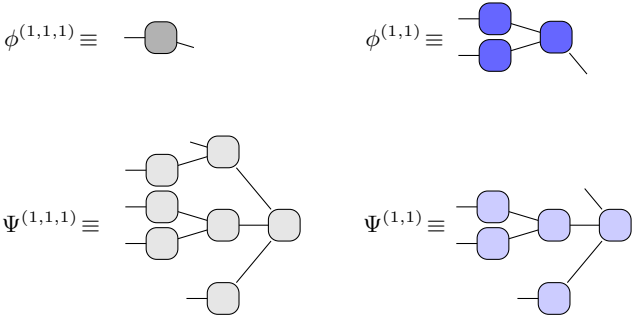}
\caption{\label{fig:spf_diagram} Diagram representing the SPFs and SHFs
associated with different nodes of the ML-MCTDH wavefunction shown
in Fig.\,\ref{fig:mctdh_wavefunction}. Here we have dropped the labels of 
the coefficient tensors in the tree tensor network.  The entire wavefunction can
be constructed by contracting together the SPF and SHF tensors for
a given node. }
\end{figure}

Working with the individual coefficient tensors and SPFs
becomes cumbersome for large tree structures. In order to simplify
expressions it is useful to define the ``single-hole functions'' (SHFs),
$\Psi_{i}^{z}$, associated with each node, $z$, by
\begin{equation}
\left|\Psi(t)\right\rangle =\sum_{i}^{N^{z_{l}}}\left|\phi_{\boldsymbol{}i}^{z_{l}}(t)\right\rangle \otimes\ket{\Psi_{i}^{z_l}}.
\end{equation}
The definitions of the SPFs and SHFs are illustrated graphically
in Fig.\,\ref{fig:spf_diagram}. In order to obtain a recursive definition 
for the SHFs that is consistent at all layers, including the first layer of the tree, we will define SHFs for the
root node as the scalar function $\Psi_{1}^{1}(t)=1$ that acts on no physical degrees
of freedom.   Further, we will append an additional index $i = 1$ to the root node coefficient tensor such that $A^1_{1_1\dots i_{d^1} i} \equiv A^1_{1_1\dots i_{d^1}}$, and define the number of SPFs at the root nodes as $N^1 = 1$.
Using this notation, the SHFs of any node $(z_l, k)$ can be obtained
recursively in terms of the coefficient tensor and
SHFs of its parent node, $z_{l}$, and the SPFs of the sibling nodes
of node $(z_{l},k)$ (that is all nodes $(z_{l},j)$ with $j\neq k$)
by
\begin{equation}
\begin{aligned}\Big|\Psi_{n}^{(z_{l},k)}(t) &\Big\rangle =  \!\sum_{i_{1}}^{n_{1}^{z_{l}}}\!\dots\!\sum_{i_{k-1}}^{n_{k-1}^{z_{l}}}\!\sum_{i_{k+1}}^{n_{k+1}^{z_{l}}}\!\dots\!\sum_{i_{d^{z_{l}}}}^{n_{d^{z_{l}}}^{z_{l}}}\sum_{m=1}^{N^{z_{l}}}\left|\Psi_{m}^{z_{l}}(t) \right\rangle \cdot \\
 & A_{i_{1}\dots i_{k-1}ni_{k+1}\dots i_{d^{z_{l}}}m}^{z_{l}}\left(\bigotimes_{\substack{j=1\\
{j\neq k}
}
}^{d^{z_{l}}}\left|\phi_{i_{j}}^{(z_{l},j)}(t)\right\rangle \right).
\end{aligned}
\label{eq:shf_recursive_definition}
\end{equation}

While the SPFs are required to be orthonormal, $\braket{\phi_{\boldsymbol{}i}^{z_{l}}(t)}{\phi_{\boldsymbol{}j}^{z_{l}}(t)}=\delta_{ij}\ \forall\ z_{l}$,
this is not the case for the SHFs constructed using Eq.\,\ref{eq:shf_recursive_definition}. Instead, we generate an overlap matrix,
$\rho_{ij}^{z_{l}}=\braket{\Psi_{i}^{z_{l}}(t)}{\Psi_{j}^{z_{l}}(t)}$,
which is referred to as the mean-field density matrix of node $z_{l}$.
This matrix can become singular whenever the SHFs are linearly
dependent, a situation that is often encountered during the early stage
of simulations when separable or weakly entangled initial condition
are used. As will be discussed later, this leads to significant
numerical issues in the evolution of ML-MCTDH wavefunctions.

Using the recursive definition of the SPFs, we may expand
the full wavefunction as
\begin{equation}
\begin{aligned}[b]\left|\Psi(t)\right\rangle  & \!=\!\sum_{i_{1}}^{n_{1}^{z_{l}}}\!\dots\!\sum_{i_{d^{z_{l}}}}^{n_{d^{z_{l}}}^{z_{l}}}\!\sum_{i}^{N^{z_{l}}}\!A_{i_{1}\dots i_{d^{z_{l}}}i}^{z_{l}}(t)\left|\Psi_{i}^{z_{l}}(t)\right\rangle\!\otimes\!\left(\!\bigotimes_{k=1}^{d^{z_{l}}}\!\left|\phi_{i_{k}}^{(z_{l},k)}(t)\!\right\rangle \!\right)\\
 & =\!\!\!\!\!\sum_{I^{z_l}}^{n_{1}^{z_{l}}\dots n_{d^{z_{l}}}^{z_{l}}}\!\sum_{i}^{n^{z_{l}}}A_{I^{z_l}i}^{z_{l}}(t)\left|\Psi_{i}^{z_{l}}(t)\right\rangle\!\otimes\!\left|\Phi_{I^{z_l}}^{z_{l}}(t)\right\rangle ,
\end{aligned}
\label{eq:wavefunction_expansion_mctdh-states}
\end{equation}
where $I^{z_l}=(i_{1},\dots, i_{d^{z}})$ is a combined
index for all indices involved in the contractions with the children
of node $z_{l}$, and $\left|\Phi_{I^{z_l}}^{z_{l}}(t)\right\rangle $
is the direct product of the SPFs associated with the children of
node $z_{l}$, which we will refer to as the configurations of node
$z_{l}$ in the following. This corresponds to an expansion of the wavefunction in
terms of the (time-dependent) direct product of the SPFs associated
with the children of node $z_{l}$ and SHFs associated with node $z_{l}$,
where the tensor $A_{I^{z_l}i}^{z_{l}}(t)$ contains the
expansion coefficients. In what follows we will prefer to work with
a matrix representation of the states, and so we will now expand the
wavefunction in terms of the primitive basis. Doing so gives the following
expression for the elements of the wavefunction coefficient tensor
\begin{equation}
\begin{aligned}\Psi_{\alpha^{z_{l}}\beta^{z_{l}}}(t) & =\!\!\!\!\!\sum_{I^{z_l}}^{n_{1}^{z_{l}}\dots n_{d^{z_{l}}}^{z_{l}}}\!\sum_{i}^{n^{z_{l}}}A_{I^{z_l}i}^{z_{l}}(t)\Phi_{\alpha^{z_{l}}I^{z_l}}^{z_{l}}(t)\Psi_{i\beta^{z_{l}}}^{z_{l}}(t)\\
 & =\left[\boldsymbol{{\Phi}}^{z_{l}}(t)\boldsymbol{{A}}^{z_{l}}(t)\boldsymbol{{\Psi}}^{z_{l}}(t)\right]_{\alpha^{z_{l}}\beta^{z_{l}}},
\end{aligned}
\label{eq:wavefunction_expansion_mctdh_matrix_elements}
\end{equation}
where $\alpha^{z_{l}}$ and $\beta^{z_{l}}$
are sets of indices for the physical degrees of freedom of
SPFs and SHFs associated with node $z_{l}$, respectively. In order to use
this notation for all nodes, we will make use of the definition of the 
SHFs for the root nodes given above, and for the leaf nodes we will define the
configurations coefficient tensor to be $\boldsymbol{{\Phi}}^{z_{l}}(t)=\boldsymbol{{I}}$.
In the final line we have written this using a matrix representation
for all of the tensors present.

This representation is particularly useful when evaluating
the ML-MCTDH EOMs, as it directly provides access to the coefficient
tensors that we seek to evolve. However, this representations is not
unique. It is possible to insert any pair of time-dependent matrices of the correct sizes that multiply to give the identity matrix between
any of the matrices in Eq.\,\ref{eq:wavefunction_expansion_mctdh_matrix_elements},
e.g.
\begin{equation}
\begin{aligned}\Psi_{\!\alpha^{z_{l}}\!\beta^{z_{l}}}\!(t)
\!=&\!\Big[\boldsymbol{{\Phi}}^{z_{l}}(t) \boldsymbol{{X}}(t)\boldsymbol{{X}}^{\text{{-}1}}(t)\boldsymbol{{A}}^{z_{l}}(t)
 \boldsymbol{{Y}}(t)\boldsymbol{{Y}}^{\text{{-}1}}(t)\boldsymbol{{\Psi}}^{z_{l}}(t)\Big]_{\!\alpha^{z_{l}}\!\beta^{z_{l}}}\\
=&\!\Big[\boldsymbol{{\Phi}}^{\prime z_{l}}(t) \boldsymbol{{A}}^{\prime z_{l}}(t)\boldsymbol{{\Psi}}^{\prime z_{l}}(t)\Big]_{\!\alpha^{z_{l}}\!\beta^{z_{l}}}.
\end{aligned}
\label{eq:wavefunction_expansion_mctdh_matrix_elements_ambiguity}
\end{equation}
In order to remove this gauge freedom, it is necessary to
specify some additional conditions, namely the static and dynamic gauge conditions.

\section{The ML-MCTDH Equations of Motion\label{sec:The-ML-MCTDH-Equations}}
In order for the ML-MCTDH approach to be efficient, it is
necessary to be able to efficiently evaluate the action of the Hamiltonian
on the ML-MCTDH wavefunction. One representation that is particularly
useful for the ML-MCTDH approach is the sum-of-product (SOP)  form
of the Hamiltonian, in which the Hamiltonian is expressed as a sum
of terms, with each term expressed as a direct product of operators
acting on the physical modes of the system. For a Hamiltonians
in this SOP form, we may write
\begin{equation}
\hat{{H}}=\sum_{r}\hat{{h}}_{r}^{z_{l}}\hat{{H}}_{r}^{z_{l}},
\end{equation}
where we have partitioned each of the product operators
into a direct product of terms $\hat{{h}}_{r}^{z_{l}}$ that act on
the physical degrees of freedom accounted for by the configurations,
and terms $\hat{{H}}_{r}^{z_{l}}$ that act on the physical degrees
of freedom accounted for by the SHF basis associated with node $z_{l}$.
For the root node, the mean-field Hamiltonians will be taken as 
the scalar $\hat{{H}}_{r}^{1}=1$.

The standard ML-MCTDH approach removes this ambiguity by
requiring that for each node the SPFs are orthonormal and remain orthonormal
through the evolution. This can be done by imposing orthonormality
of the SPFs at a given time, which amounts to a static gauge condition. However on
its own this condition is not sufficient to remove all ambiguities
in this representation. It is still possible to insert arbitrary unitary
factors (and their inverses) between terms. This ambiguity can be removed
by enforcing a dynamic gauge condition on the SPFs associated with each
of the non-root nodes $z_{l}$\citep{BECK20001,WANG2003,MANTHE2008}
\begin{equation}
\boldsymbol{{\phi}}^{z_{l}\dagger}(t)\tder{{\boldsymbol{{\phi}}}}^{z_{l}}(t)=-\boldsymbol{{i}}\boldsymbol{{x}}^{z_{l}}(t),\label{eq:spf_gauge}
\end{equation}
where $\boldsymbol{{x}}^{z_{l}}(t)$ is a Hermitian matrix
for all times that can, in general, depend on the ML-MCTDH wavefunction.
The choice of the matrix $\boldsymbol{{x}}^{z_l}(t)$ does not effect
the accuracy of the ML-MCTDH approach, however, it can influence the
numerical efficiency of the integration of the final EOMs.

As the SPFs can be expanded in terms of the coefficient
tensor and SPFs of the children of node $z_{l}$,
\begin{equation}
\boldsymbol{{\phi}}^{z_{l}}(t)=\left(\bigotimes_{k=1}^{d^{z_{l}}}\boldsymbol{{\phi}}^{(z_{l},k)}(t)\right)\boldsymbol{{A}}^{z_{l}}(t)=\boldsymbol{{\Phi}}^{z_{l}}(t)\boldsymbol{{A}}^{z_{l}}(t),
\end{equation}
it can readily be shown that these gauge conditions apply
constraints to all non-root coefficient tensors. The gauge condition
(Eq.\,\ref{eq:spf_gauge}) for each of the SPFs of the children node
$z_{l}$ leads to the following constraint on the coefficient tensors
for the configurations,
\begin{equation}
\boldsymbol{{\Phi}}^{z_{l}\dagger}(t)\tder{{\boldsymbol{{\Phi}}}}^{z_{l}}(t)=-\boldsymbol{{i}}\left(\bigoplus_{k=1}^{d^{z_{l}}}\boldsymbol{{x}}^{(z_{l},k)}(t)\right)=-\boldsymbol{{i}}\boldsymbol{{X}}^{z_{l}}(t),\label{eq:spf_gauge-1}
\end{equation}
where $\bigoplus_{k}$ denotes a direct sum over matrices
indexed by $k$. As such, the coefficient tensors are constrained
to satisfy
\begin{equation}
\boldsymbol{{A}}^{z_{l}\dagger}(t)\tder{\boldsymbol{{A}}}^{z_{l}}(t)=-i\boldsymbol{{x}}^{z_{l}}(t)+i\boldsymbol{{A}}^{z_{l}\dagger}(t)\boldsymbol{{X}}^{z_{l}}(t)\boldsymbol{{A}}^{z_{l}}(t).\label{eq:Atensor_gauge}
\end{equation}

These conditions ensure that the configurations remain orthonormal
and that the coefficient matrix remains semi-unitary for all time.
As a consequence, any non-orthonormality associated with the matrix
expansion of the wavefunction given in Eq.\,\ref{eq:wavefunction_expansion_mctdh_matrix_elements}
must be accounted for by the SHF coefficient tensor. Thus the standard
ML-MCTDH approach expands the wavefunction at each (non-root) node
in terms of a direct product of orthonormal SPFs and non-orthonormal,
and in some cases linearly dependent, SHFs with an expansion coefficient
matrix that is semi-unitary.

With these constraints and the specification of the Hamiltonian
we are now in a position to derive the ML-MCTDH
EOMs. The standard derivation of the ML-MCTDH EOMs has been presented
in a number of places,\citep{WANG2003,MANTHE2008} here we will present a slightly different path
to the final ML-MCTDH EOMs. In the derivation that follows,
all matrices and states are time-dependent. However, in order to prevent
the equations becoming too cumbersome, we will not indicate this explicitly
in intermediate equations. As is standard, we begin by applying the Dirac-Frenkel
variational principle\citep{WANG2003,MANTHE2008}
\begin{equation}
\big\langle\delta\Psi\big|i\frac{{\partial}}{\partial t}-\hat{{H}}\big|\Psi\big\rangle=0
\end{equation}
to the ML-MCTDH wavefunction.  The variation of the wavefunction
can be written in the form\citep{MANTHE2008}
\begin{equation}
\big|\delta\Psi\big\rangle=\sum_{z_{l}\in\Psi}\sum_{I^{z_l}}\sum_{i^{z_{l}}}\delta A_{I^{z_{l}}i^{z_{l}}}^{z_{l}}\big|\Phi_{I^{z_l}}^{z_{l}}\big\rangle\otimes\big|{\Psi_{i^{z_{l}}}^{z_{l}}}\big\rangle.
\end{equation}
Inserting this into the Dirac-Frenkel variational principle,
we have
\begin{equation}
\sum_{z_{l}\in\Psi}\sum_{I^{z_{l}}}\sum_{i^{z_{l}}}\delta A_{I^{z_l} i^{z_{l}}}^{z_{l}\dagger}\big\langle\Phi_{I^{z_{l}}}^{z_{l}}\big|\otimes\big\langle\Psi_{i^{z_{l}}}^{z_{l}}\big|i\frac{{\partial}}{\partial t}-\hat{{H}}\ket{\Psi}=0.
\end{equation}
The variations of the coefficient tensors $\delta A_{I^{z_{l}}i^{z_l}}^{z_{l}\dagger}$ are
independent, and we can therefore consider variations of each of the
coefficient tensors independently. Doing so simplifies this expression
to give
\begin{equation}
\sum_{I^{z_l}}\sum_{i^{z_l}}\delta A_{I^{z_l}i^{z_l}}^{z_{l}\dagger}\big\langle\Phi_{I^{z_l}}^{z_{l}}\big|\otimes\big\langle\Psi_{i^{z_l}}^{z_{l}}\big|i\frac{{\partial}}{\partial t}-\hat{{H}}\big|\Psi\big\rangle=0,\label{eq:variations_single_site}
\end{equation}
where now we are only considering variations of the coefficient
tensor of node $z_{l}$. We can now expand the time derivative
of the wavefunction using the representation of the wavefunction given
in Eq.\,\ref{eq:wavefunction_expansion_mctdh-states}, which yields
\begin{equation}
\begin{aligned}\ket{\dot{\Psi}}= & \sum_{J^{z_l}}\sum_{j}\Big(\dot{A}{}_{J^{z_l}j}^{z_{l}}\ket{\Phi_{J^{z_l}}^{z_{l}}}\ket{\Psi_{j}^{z_{l}}}\\
+ & A_{J^{z_l}j}^{z_{l}}\big|\dot{{\Phi}}_{J^{z_l}}^{z_{l}}\big\rangle\big|\Psi_{j}^{z_{l}}\big\rangle+A_{J^{z_l}j}^{z_{l}}\big|\Phi_{J^{z_l}}^{z_{l}}\big\rangle\big|\dot{\Psi}_{j_{}}^{z_{l}}\big\rangle\Big).
\end{aligned}
\end{equation}

Inserting this expression into Eq.\,\ref{eq:variations_single_site},
and noting that the resultant expression holds for all allowed variations
of the coefficient tensor, we may equate coefficients to remove the
dependence on the variations. This yields an equation that can be written in terms of the
configurations and SHF and node coefficient matrices as
\begin{equation}
\begin{aligned}\boldsymbol{\Phi}^{z_{l}\dagger}\tder{{\boldsymbol{{\Phi}}}}^{z_{l}}\boldsymbol{{A}}^{z_{l}}\boldsymbol{{\Psi}}^{z_{l}}\boldsymbol{{\Psi}}^{z_{l}\dagger}+\boldsymbol{\Phi}^{z_{l}\dagger}\boldsymbol{\Phi}^{z_{l}}\tder{{\boldsymbol{{A}}}}^{z_{l}}\boldsymbol{{\Psi}}^{z_{l}}\boldsymbol{{\Psi}}^{z_{l}\dagger}\\
+\boldsymbol{\Phi}^{z_{l}\dagger}\boldsymbol{\Phi}^{z_{l}}\boldsymbol{{A}}^{z_{l}}\tder{{\boldsymbol{{\Psi}}}}^{z_{l}}\boldsymbol{{\Psi}}^{z_{l}\dagger}=-\frac{{i}}{\hbar}\sum_{r}\boldsymbol{{h}}_{r}^{z_{l}}\boldsymbol{{A}}^{z_{l}} & \boldsymbol{{H}}_{r}^{z_{l}}.
\end{aligned}
\label{eq:mlmctdh_variations-1}
\end{equation}
Here we have used the sum-of-product representation of the
Hamiltonian, and we have introduced the SPF, $\boldsymbol{{h}}_{r}^{z_{l}}$(t),
and mean-field, $\boldsymbol{{H}}_{r}^{z_{l}}(t)$, Hamiltonian matrices, respectively.
These Hamiltonians have the matrix elements
\begin{align}
\left[\boldsymbol{{h}}_{r}^{z_{l}}(t)\right]_{IJ} & =\bra{\Phi_{I}^{z_{l}}(t)}\hat{h}_{r}^{z_{l}}\ket{\Phi_{J}^{z_{l}}(t)},\\
\left[\boldsymbol{{H}}_{r}^{z_{l}}(t)\right]_{ij} & =\bra{\Psi_{i}^{z_{l}}(t)}\hat{H}_{r}^{z_{l}}\ket{\Psi_{j}^{z_{l}}(t)}.
\end{align}
The SPF and mean-field Hamiltonian matrices depend on all
coefficient tensors used to evaluate the configurations and SHFs for
node $z_{l}$, respectively. Here for notational simplicity we will
not explicitly indicate this dependence.

For the root node, we can apply the gauge conditions for
the SPFs, use the fact that the SPF coefficient tensor is semi-unitary, and
make use of the fact that we have defined the SHFs as a time-independent scalar to
obtain\citep{BECK20001,WANG2003,MANTHE2008} 
\begin{equation}
\tder{{\boldsymbol{{A}}}}^{1}(t)=-i\left(\frac{1}{\hbar}\boldsymbol{{h}}{}^{1}(t)-\boldsymbol{{X}}^{1}(t)\right)\boldsymbol{{A}}^1(t).\label{eq:root-eom-mlmctdh}
\end{equation}

For a given non-root node $z_{l}$, applying the gauge
conditions for the SPFs simplifies Eq.\,\ref{eq:mlmctdh_variations-1}, giving
\begin{equation}
\begin{aligned}-\boldsymbol{{X}}^{z_{l}}(t)\boldsymbol{{A}}^{z_{l}}\boldsymbol{{\rho}}^{z_{l}}+\tder{\boldsymbol{A}}^{z_{l}}\boldsymbol{{\rho}}^{z_{l}}\\
+\boldsymbol{{A}}^{z_{l}}\dot{{\boldsymbol{{\Psi}}}}^{z_{l}}\boldsymbol{{\Psi}}^{z_{l}\dagger}=-\frac{{i}}{\hbar} & \sum_{r}\boldsymbol{{h}}_{r}^{z_{l}}\boldsymbol{{A}}^{z_{l}}\boldsymbol{{H}}_{r}^{z_{l}},
\end{aligned}
\label{eq:mlmctdh_variations-1-1}
\end{equation}
where $\boldsymbol{{\rho}}^{z_{l}}=\boldsymbol{\Psi}^{z_{l}}\boldsymbol{\Psi}^{z_{l}\dagger}$
is the mean-field density matrix discussed previously, which like
the mean-field Hamiltonian depends on the coefficient tensors of all
nodes used in the evaluation of the SHFs. This equation describes
the evolution of both the coefficient tensor associated with a node
as well as the SHF coefficient tensors, and thus all of
the coefficient tensors of nodes used to define the SHF coefficient
tensors.

It is possible to arrive at the EOMs for the coefficient
tensor of node $z_l$ by expanding the derivative of the SHF coefficient
tensor using the recursive expression for the SHF coefficient tensors,
as is the standard approach taken in arriving at the ML-MCTDH EOMs.\citep{WANG2003} 
However, this is a relatively cumbersome task.
An alternative and considerably faster approach simply involves applying
the projector
\begin{equation}
\boldsymbol{{Q}}^{z_{l}}=\boldsymbol{{I}}-\boldsymbol{{A}}^{z_{l}}\boldsymbol{{A}}^{z_{l}\dagger}=\boldsymbol{I}-\boldsymbol{P}^{z_{l}},\label{eq:projector}
\end{equation}
where $\boldsymbol{{I}}$ is the identity matrix, to Eq.
\ref{eq:mlmctdh_variations-1-1} from the left. This projector acts
on the matrix $\boldsymbol{{A}}^{z_{l}}$ to give
\begin{equation}
\boldsymbol{{Q}}^{z_{l}}\boldsymbol{{A}}^{z_{l}}=\boldsymbol{A}^{z_{l}}-\boldsymbol{{A}}^{z_{l}}\boldsymbol{{A}}^{z_{l}\dagger}\boldsymbol{{A}}^{z_{l}}=\boldsymbol{{0}},\label{eq:projector2}
\end{equation}
due to the semi-unitarity of the coefficient tensors. As
such, the term containing the derivative of the SHF vanishes when
this projector is applied. Using these facts we find
\begin{equation}
\begin{aligned}-\boldsymbol{{Q}}^{z_{l}}\boldsymbol{{X}}^{z_{l}}(t)\boldsymbol{{A}}^{z_{l}}\boldsymbol{{\rho}}^{z_{l}}+\tder{{\boldsymbol{{A}}}}^{z_{l}}\boldsymbol{{\rho}}^{z_{l}}\\
-\boldsymbol{{A}}^{z_{l}}\boldsymbol{{A}}^{z_{l}\dagger}\tder{{\boldsymbol{{A}}}}^{z_{l}}\boldsymbol{{\rho}}^{z_{l}}=-\frac{{i}}{\hbar}\boldsymbol{{Q}}^{z_{l}} & \sum_{r}\boldsymbol{{h}}_{r}^{z_{l}}\boldsymbol{{A}}^{z_{l}}\boldsymbol{{H}}_{r}^{z_{l}}.
\end{aligned}
\label{eq:mlmctdh_variations-1-1-1}
\end{equation}
Inserting Eq.~\ref{eq:Atensor_gauge} and rearranging terms gives
the final ML-MCTDH EOMs for all non-root nodes\citep{WANG2003,MANTHE2008}
\begin{equation}
\begin{aligned}\tder{{\boldsymbol{{A}}}}^{z_{l}}(t)\!= \!& -\frac{{i}}{\hbar}\boldsymbol{{Q}}^{z_{l}}(t)\!\sum_{r}\!\boldsymbol{{h}}_{r}^{z_{l}}(t)\boldsymbol{{A}}^{z_{l}}(t)\boldsymbol{{H}}_{r}^{z_{l}}(t)\boldsymbol{{\rho}}^{z_{l}}(t)^{-1}\\
 & -i\boldsymbol{{A}}^{z_{l}}(t)\boldsymbol{{x}}^{z_{l}}(t)+i\boldsymbol{{X}}^{z_{l}}(t)\boldsymbol{{A}}^{z_{l}}(t).
\end{aligned}
\label{eq:mlmctdh_eom}
\end{equation}
This expression contains the inverse of the mean-field density matrix
(the SHF overlap matrix). Whenever the SHFs are linearly dependent,
the mean-field density matrix becomes singular and its inverse ill-defined.
This in turn leads to ill-defined ML-MCTDH EOMs, as they require the
evaluation and application of an unbounded operator. When the SHFs are nearly
linearly dependent, the condition number of the mean-field density
matrix becomes large, and the ML-MCTDH EOMs for the SPFs can become
very stiff, significantly increasing the computation effort required
for integration. In such cases, the ML-MCTDH EOMs can lead to rapid evolution of the 
coefficient tensors, giving rise to a rapid evolution of the weakly occupied SPFs.\citep{MEYER2018149} 
In practice it is necessary to regularize the inverse
of the mean-field density matrix when evaluating the EOMs. A number
of different strategies have been developed to regularize these equations,
however in all cases they introduce a bias into the equations, and
it becomes necessary to converge the dynamics with respect to this
bias. For many problems, moderate levels of regularization do not
significantly alter the dynamics and it is possible to obtain dynamics
that are converged with respect to the regularization parameter. However,
for some problems small regularization parameters may be required
to obtain accurate dynamics. As the inverse of the standard regularization
parameter is directly related to the condition number of the regularized
inverse of the mean-field density matrix, small regularization parameters
correspond to increased computation effort. In severe cases, this
may even prevent converged dynamics from being obtained.

Before one can solve the ML-MCTDH EOMs, it is necessary
to specify the constraint matrices. The standard choice for this constraint
matrix is\citep{WANG2003,BECK20001,MANTHE2008}
\begin{equation}
\boldsymbol{{x}}^{z_{l}}(t)=\boldsymbol{0}.\label{eq:standard_gauge}
\end{equation}
This choice minimises the motion of configurations, and
is particularly useful when using constant mean-field integration
(CMF) schemes discussed in Sec. I.A of the Supplementary Information. 
Alternative choices for the dynamic gauge condition may lead to numerical benefits in some situations.\citep{WANG2003,BECK20001,MANTHE2008}  

\section{Singularity-Free Equations of Motion\label{sec:Singularity-Free-Equations}}
As discussed in the previous section, the singularities present in
the ML-MCTDH EOMs arise due to the use of a non-orthonormal set of
SHFs in the expansion of the wavefunction. It is therefore reasonable
to ask whether the use of orthonormalized SHFs will lead to non-singular
EOMs. Here we will demonstrate that such a choice leads
to a family of singularity-free EOMs that differ in the
choice of their gauge conditions, and demonstrate that the PSI EOMs arise
from one such gauge choice. In order to explore this we consider an
alternative expansion of the full wavefunction than that given by
Eq.\,\ref{eq:wavefunction_expansion_mctdh_matrix_elements}, one in
which the SHF basis functions are orthonormal.

To construct such a representation, we will make use of
the following decomposition of the SHF coefficient matrix for the
non-root node $z_{l}$
\begin{equation}
\boldsymbol{{\Psi}}^{z_{l}}(t)=\boldsymbol{{R}}^{z_{l}}(t)\orthrep{{\boldsymbol{{\Psi}}}}^{z_{l}}(t),\label{eq:SHF_decomposition}
\end{equation}
where $\orthrep{{\boldsymbol{{\Psi}}}}^{z_{l}}$ is a
semi-unitary matrix defining an orthonormal set of SHFs represented
in terms of the primitive basis, and $\boldsymbol{{R}}^{z_{l}}$
contains the expansion coefficients of the old SHF in terms of these
new functions. While a decomposition of this form can be performed for any
matrix, this will in general be impractical for the full
expanded SHF coefficient matrix due to its large size. In practice,
a recursive approach is used for constructing this decomposition that
makes use of the definition of the SHFs given in Eq.\,\ref{eq:shf_recursive_definition}.
Such a strategy allows for the construction of a tree tensor network
representation of the transformed SHF coefficient tensor, which is
necessary for practical applications of the approach.

The wavefunction can be expanded in terms of the standard
configurations and the new SHF states at node $z_{l}$ as
\begin{equation}
\begin{aligned}\left|\Psi(t)\right\rangle  & =\!\!\!\!\!\!\sum_{I^{z_l}}^{n_{1}^{z_{l}}\dots n_{d^{z_{l}}}^{z_{l}}}\!\sum_{i,j}^{n^{z_{l}}}\!A_{I^{z_l}i}^{z_{l}}(t)\left|\Phi_{I^{z_l}}^{z_{l}}(t)\right\rangle\!\otimes\! R_{ij}^{z_{l}}(t)\!\left|\orthrep{\Psi}_{j}^{z_{l}}(t)\!\right\rangle \\
 & =\!\!\!\!\!\!\sum_{I^{z_l}}^{n_{1}^{z_{l}}\dots n_{d^{z_{l}}}^{z_{l}}}\!\sum_{j}^{n^{z_{l}}}\orthrep{A}_{I^{z_l}j}^{z_{l}}(t)\left|\Phi_{I^{z_l}}^{z_{l}}(t)\right\rangle \!\otimes\!\left|\orthrep{\Psi}_{j}^{z_{l}}(t)\!\right\rangle .
\end{aligned}
\label{eq:transformed_wavefunction_representation}
\end{equation}
Here we have introduced the transformed coefficient tensors
$\orthrep{A}_{\boldsymbol{{I}}j}^{z_{l}}(t)$ which are related to the standard coefficient tensors by the matrix expression

\begin{equation}
\orthrep{{\boldsymbol{{A}}}}^{z_{l}}(t)=\boldsymbol{{A}}^{z_{l}}(t)\boldsymbol{{R}}^{z_{l}}(t).\label{eq:transformed_Atilde_definition}
\end{equation}
We can also express the wavefunction in terms
of the new SHF states and the original SPF states for the non-root
node $z_{l}$ as
\begin{equation}
\begin{aligned}\left|\Psi(t)\right\rangle  & =\sum_{i,j}^{n^{z_{l}}}R_{ij}^{z_{l}}(t)\big|\phi_{i}^{z_{l}}(t)\big\rangle \otimes\big|\orthrep{\Psi}_{j}^{z_{l}}(t)\big\rangle . \end{aligned}
\end{equation}
In contrast to Eq.\,\ref{eq:wavefunction_expansion_mctdh-states},
both of the wavefunction expressions immediately above correspond to an expansion in terms
of an incomplete but orthonormal set of functions. For the root node,
the original representation already corresponds to an expansion of
the wavefunction in terms of an orthonormal set of functions as there
are no SHFs associated with the root. As such, we use the standard
representation, and for simplicity we will define $\orthrep{{\boldsymbol{{A}}}}^{1}(t)\equiv\boldsymbol{{A}}^{1}(t).$
Note we have not introduced a set of transformed SHFs
for the root node or for a $\boldsymbol{{R}}^{1}(t)$ matrix.

When the original SHFs are linearly
independent, the transformed SHFs are an orthonormal set of functions
that span the same space for all non-root nodes. When the original SHFs are linearly dependent,
this is no longer the case. The transformed SHFs are a set of orthonormal
functions that span the union of the space spanned by the original
SHFs, as well as an arbitrarily selected space that is orthonormal
to this. The arbitrarily selected functions do not contribute to
the overall wavefunction, which is the reason they can be chosen freely.

Regardless of whether the original SHFs are linearly dependent
or not, Eq.\,\ref{eq:SHF_decomposition} is not unique. It is possible
to insert any unitary matrix and its adjoint between the two terms
without changing the value of the SHF coefficient tensor. This additional
gauge freedom relative to the ML-MCTDH representation requires
us to specify an additional gauge condition to uniquely define the
EOMs. Here, we will do this by applying a condition for $\orthrep{{\boldsymbol{{\Psi}}}}^{z_{l}}(t)$
\begin{equation}
\dotorthrep{\boldsymbol{\Psi}}^{z_{l}}(t)\orthrep{\boldsymbol{\Psi}}^{z_{l}\dagger}(t)=-\boldsymbol{{i}}\boldsymbol{{y}}^{z_{l}}(t),\label{eq:shf_gauge}
\end{equation}
where $\boldsymbol{{y}}^{z_{l}}(t)$ is Hermitian, which
ensures that the transformed SHFs remain orthonormal. Another gauge choice
that is of particular interest is to set the constraint matrices for
the SPFs at each non-root node to $\boldsymbol{x}^{z_{l}}(t)=\boldsymbol{0}$
and the constraint matrix for the transformed SHFs at each non-root
node to $\boldsymbol{y}^{z_{l}}(t)=\boldsymbol{0}$. This choice gives
rise to the PSI EOM.

The dynamics of a ML-MCTDH wavefunction is described by
the ML-MCTDH EOMs for the standard coefficient tensors given by Eqs.\,\ref{eq:root-eom-mlmctdh} and \ref{eq:mlmctdh_eom}. Expressing the
ML-MCTDH EOM for the root node in terms of the transformed coefficient
tensors, we obtain the standard result
\begin{equation}
\dotorthrep{\boldsymbol{A}}^{1}(t)=\tder{\boldsymbol{A}}^{1}(t)=-i\left(\frac{1}{\hbar}\boldsymbol{{h}}{}^{1}(t)-\boldsymbol{{X}}^{1}(t)\right)\orthrep{\boldsymbol{A}}^1(t).\label{eq:alt_rep_eom_root}
\end{equation}
For all non-root nodes, we obtain
\begin{equation}
\begin{aligned}\tder{\boldsymbol{A}}^{z_{l}}(t)\!=\! & -\!\frac{i}{\hbar}\boldsymbol{Q}^{z_{l}}(t)\!\sum_{r}\!\boldsymbol{h}_{r}^{z_{l}}(t)\orthrep{\boldsymbol{A}}^{z_{l}}(t)\orthrep{\boldsymbol{H}}_{r}^{z_{l}}(t)\!\left(\!\boldsymbol{R}^{z_{l}}(t)\right)^{\text{-}1}\\
 & -i\boldsymbol{A}^{z_{l}}(t)\boldsymbol{x}^{z_{l}}(t)+i\boldsymbol{X}^{z_{l}}(t)\boldsymbol{A}^{z_{l}}(t),
\end{aligned}
\label{eq:alt_rep_eom}
\end{equation}
where we have introduced the transformed mean-field Hamiltonian
matrices, $\orthrep{\boldsymbol{H}}_{r}^{z_{l}}(t)$, that each
have matrix elements
\begin{equation}
\left[\orthrep{\boldsymbol{H}}_{r}^{z_{l}}(t)\right]_{ij}=\bra{\orthrep{\Psi}_{i}^{z_{l}}(t)}\hat{H}_{r}^{z_{l}}\ket{\orthrep{\Psi}_{j}^{z_{l}}(t)}.
\end{equation}
Whenever the mean-field density matrix is singular, so too
is the matrix $\boldsymbol{R}^{z_{l}}(t)$. Thus we have not made
any practical gains at this stage. In order to make progress, we need to consider an
alternative approach for treating the dynamics of the standard coefficient
tensors. To do this, we will start by considering the derivative of
the transformed coefficient tensor, $\orthrep{\boldsymbol{A}}^{z_{l}}(t)$.
Upon rearranging we find that
\begin{equation}
\tder{\boldsymbol{A}}^{z_{l}}\boldsymbol{R}^{z_{l}}=\dotorthrep{\boldsymbol{A}}^{z_{l}}-\boldsymbol{A}^{z_{l}}\tder{\boldsymbol{R}}^{z_{l}}.\label{eq:atilde_derivative}
\end{equation}
The EOM for the standard coefficient tensors, $\boldsymbol{A}^{z_{l}}$,
that describe the dynamics of the full ML-MCTDH wavefunction can be
written in terms of EOMs for the transformed coefficient tensor, $\orthrep{\boldsymbol{A}}^{z_{l}}$,
and the EOM for the $\boldsymbol{R}^{z_{l}}$ tensors. We will
now obtain EOMs for these two objects.

In order to obtain EOMs for these transformed coefficient
tensors, we begin by rewriting Eq.\,\ref{eq:mlmctdh_variations-1}
in terms of the transformed coefficient tensors and SHF matrices.
This yields
\begin{equation}
\begin{aligned}\boldsymbol{\Phi}^{z_{l}\dagger}\tder{{\boldsymbol{{\Phi}}}}^{z_{l}}\orthrep{{\boldsymbol{{A}}}}^{z_{l}}\orthrep{{\boldsymbol{{\Psi}}}}^{z_{l}}\orthrep{{\boldsymbol{{\Psi}}}}^{z_{l}\dagger}\boldsymbol{R}^{z_{l}}+\boldsymbol{\Phi}^{z_{l}\dagger}\boldsymbol{\Phi}^{z_{l}}\dotorthrep{{\boldsymbol{{A}}}}^{z_{l}}\orthrep{{\boldsymbol{{\Psi}}}}^{z_{l}}\orthrep{{\boldsymbol{{\Psi}}}}^{z_{l}\dagger}\boldsymbol{R}^{z_{l}}\\
+\boldsymbol{\Phi}^{z_{l}\dagger}\boldsymbol{\Phi}^{z_{l}}\orthrep{{\boldsymbol{{A}}}}^{z_{l}}\dotorthrep{{\boldsymbol{{\Psi}}}}^{z_{l}}\orthrep{{\boldsymbol{{\Psi}}}}^{z_{l}\dagger}\boldsymbol{R}^{z_{l}}=-\frac{{i}}{\hbar}\sum_{r}\boldsymbol{{h}}_{r}^{z_{l}}\orthrep{{\boldsymbol{{A}}}}^{z_{l}}\orthrep{{\boldsymbol{{H}}}}_{r}^{z_{l}} & \boldsymbol{R}^{z_{l}}.
\end{aligned}
\label{eq:psi_eom_1_full_expansion}
\end{equation}
Applying the gauge conditions (Eqs.\,\ref{eq:spf_gauge-1}
and \ref{eq:shf_gauge} and semi-unitarity) for the coefficient tensors and transformed SHF tensors yields an EOM
for $\dotorthrep{\boldsymbol{A}}^{z_{l}}$
\begin{equation}
\begin{aligned}\dotorthrep{{\boldsymbol{{A}}}}^{z_{l}}(t)=-\frac{{i}}{\hbar}\sum_{r} & \boldsymbol{{h}}_{r}^{z_{l}}(t)\orthrep{{\boldsymbol{{A}}}}^{z_{l}}(t)\orthrep{{\boldsymbol{{H}}}}_{r}^{z_{l}}(t)\\
 & +i\boldsymbol{{X}}^{z_{l}}(t)\orthrep{{\boldsymbol{{A}}}}^{z_{l}}(t)+i\orthrep{{\boldsymbol{{A}}}}^{z_{l}}(t)\boldsymbol{{y}}^{z_{l}}(t).
\end{aligned}
\label{eq:sf_eom_1}
\end{equation}
These EOMs are non-singular regardless of whether or not
the original mean-field density matrix is singular, and correspond
to Eq.\,\ref{eq:mlmctdh_variations-1-1} expressed in terms
of the transformed coefficient tensors. As such, Eq.\,\ref{eq:sf_eom_1}
accounts for the evolution of both the standard coefficient tensor
 and the standard SHF coefficient tensor associated with node $z_{l}$.
This equation has the same form as the EOM for the root node, and
so simply correspond to evolution under the full Hamiltonian evaluated
in the time-dependent transformed basis associated with node $z_{l}$.

We now need to obtain an EOM for the $\boldsymbol{R}^{z_{l}}$
tensors. This can be done in a number of different ways. One possible
approach is to differentiate Eq.\,\ref{eq:SHF_decomposition}, and
make use of the known evolution  equation for the standard SHFs. 
However, an easier approach is obtained by expanding the derivative
of $\orthrep{\boldsymbol{A}}^{z_{l}}$ in Eq.\,\ref{eq:sf_eom_1}.
Doing so we obtain
\begin{equation}
\begin{aligned}\dot{{\boldsymbol{{A}}}}^{z_{l}}\boldsymbol{{R}}^{z_{l}}+\boldsymbol{{A}}^{z_{l}}\dot{{\boldsymbol{{R}}}}{}^{z_{l}}= & -\frac{{i}}{\hbar}\sum_{r}\boldsymbol{{h}}_{r}^{z_{l}}\boldsymbol{{A}}^{z_{l}}\boldsymbol{{R}}^{z_{l}}\orthrep{{\boldsymbol{{H}}}}_{r}^{z_{l}}\\
+i\boldsymbol{{X}}^{z_{l}} & \boldsymbol{{A}}^{z_{l}}\boldsymbol{{R}}^{z_{l}}+i\boldsymbol{{A}}^{z_{l}}\boldsymbol{{R}}^{z_{l}}\boldsymbol{{y}^{z_{l}}} ,\label{eq:26_rcoeff}
\end{aligned}
\end{equation}
which is just Eq.\,\ref{eq:mlmctdh_variations-1-1} using the decomposition of the
SHFs given in Eq.\,\ref{eq:SHF_decomposition} and the new dynamic gauge
conditions for the transformed SHFs. In the derivation of the ML-MCTDH EOMs presented in
the previous section, the application of the projector $\boldsymbol{Q}^{z_{l}}$
defined in Eq.\,\ref{eq:projector} to Eq.\,\ref{eq:mlmctdh_variations-1-1}
leads to the EOMs for the standard coefficient tensors.
This process removes the contributions associated with the evolution
of the SHFs, while retaining the contributions associated with the
derivative of the coefficient tensors. Here we are interested
in the term excluded by that projector, and so we apply the complementary projection operator to
$\boldsymbol{Q}^{z_{l}}$, namely the projector $\boldsymbol{{P}}^{z_{l}}=\boldsymbol{{A}}^{z_{l}}\boldsymbol{{A}}^{z_{l}\dagger},\label{eq:Qprojector}$
to Eq.\,\ref{eq:26_rcoeff}. After using $\boldsymbol{{P}}^{z_{l}}$, rearranging, and inserting
Eq.\,\ref{eq:Atensor_gauge}, we arrive at
\begin{equation}
\begin{aligned}\boldsymbol{{A}}^{z_{l}}\tder{{\boldsymbol{{R}}}}{z_{l}}\!=\!\boldsymbol{{A}}^{z_{l}}\Big(\! - &\frac{{i}}{\hbar}\sum_{r}\!\boldsymbol{{A}}^{z_{l}\dagger}\boldsymbol{{h}}_{r}^{z_{l}}\boldsymbol{{A}}^{z_{l}}\boldsymbol{{R}}^{z_{l}}\orthrep{{\boldsymbol{{H}}}}_{r}^{z_{l}}\\
 & +i\boldsymbol{{x}}^{z_{l}}\boldsymbol{{R}}^{z_{l}}+i\boldsymbol{{R}}^{z_{l}}\boldsymbol{{y}}^{z_{l}}\Big).
\end{aligned}
\end{equation}
Finally, the EOMs for $\boldsymbol{{R}}^{z}$ are obtained by
applying $\boldsymbol{{A}}^{z\dagger}$ from the left,
\begin{equation}
\begin{aligned}\tder{{\boldsymbol{{R}}}}^{z_{l}}(t)= & -\frac{{i}}{\hbar}\sum_{r}\boldsymbol{{A}}^{z_{l}\dagger}(t)\boldsymbol{{h}}_{r}^{z_{l}}(t)\boldsymbol{{A}}^{z_{l}}(t)\boldsymbol{{R}}^{z_{l}}(t)\orthrep{{\boldsymbol{{H}}}}_{r}^{z_{l}}(t)\\
 & +i\boldsymbol{{x}}^{z_{l}}(t)\boldsymbol{{R}}^{z_{l}}(t)+i\boldsymbol{{R}}^{z_{l}}(t)\boldsymbol{{y}}^{z_{l}}(t).
\end{aligned}
\label{eq:sf_eom_2}
\end{equation}
As long as the gauge conditions chosen are free of singularities,
this equation, like Eq.\,\ref{eq:sf_eom_1}, will only involve bounded matrices regardless of whether or not the mean-field density matrix is singular.

Eqs.\,\ref{eq:sf_eom_1} and \ref{eq:sf_eom_2} account
for the evolution of the transformed coefficient tensors $\orthrep{\boldsymbol{A}}^{z_{l}}$
and the $\boldsymbol{{R}}^{z}$ matrices, respectively. Both of these
matrices arise as the expansion coefficients of the full wavefunction
in terms of different sets of orthonormal functions. In arriving
at these equations, no explicit constraints have been placed on the
evolution of the coefficient tensors, only the basis states have
been constrained. In terms of the standard ML-MCTDH representation
of the wavefunction, the evolution of the transformed coefficient
tensors corresponds to evolution of the standard coefficient tensor,
$\boldsymbol{A}^{z_{l}}$, and the SHF coefficient tensor $\boldsymbol{\Psi}^{z_{l}}$.
The evolution of the $\boldsymbol{{R}}^{z}$ matrices corresponds
to evolution of the SHF coefficient tensors. Neither of these equations
solely evolve the standard coefficient tensors, and therefore cannot
be directly applied to evolve all nodes of the ML-MCTDH wavefunction.
However, it is possible to construct integration schemes that make
use of these equations. In order to do this, it is
necessary to linearize the ML-MCTDH EOMs given by Eqs.\,\ref{eq:alt_rep_eom_root}
and \ref{eq:alt_rep_eom} and make use of a Trotter splitting of the resultant
propagators. This process is outlined in Sec. I.B of the Supplementary Information.
Importantly, this process shows that, as in the case of the CMF and linearized
forms of the standard ML-MCTDH EOMs, the equations for updating the
coefficient tensors at each node are independent, and as such this
process can be performed in parallel. We will leave discussion of
the practical implementation of parallel integration schemes and their
numerical performance to future work. For simplicity, in Sec. I.B of the Supplementary Information 
we have considered
the specific choice for the dynamic gauge conditions used by the PSI, namely
\begin{equation}
\boldsymbol{{x}}^{z_{l}}(t)=\boldsymbol{{0}}\ \ \ \ \text{{and}}\ \ \ \boldsymbol{{y}}^{z_{l}}(t)=\boldsymbol{{0}}.\label{eq:psi_gauge_condition}
\end{equation}
We discuss and present numerical results obtained using a potentially useful 
alternative choice for the dynamic gauge conditions in Sec. III of the 
Supplementary Information.

Inserting the gauge conditions given in Eq.\,\ref{eq:psi_gauge_condition}
into Eqs.\,\ref{eq:sf_eom_1} and \ref{eq:sf_eom_2} yields
\begin{equation}
\begin{aligned}\dotorthrep{{\boldsymbol{{A}}}}^{z_{l}}(t)= & -\frac{{i}}{\hbar}\sum_{r}\boldsymbol{{h}}_{r}^{z_{l}}(t)\orthrep{{\boldsymbol{{A}}}}^{z_{l}}(t)\orthrep{{\boldsymbol{{H}}}}_{r}^{z_{l}}(t),\end{aligned}
\label{eq:psi_eom_1}
\end{equation}
and
\begin{equation}
\tder{{\boldsymbol{{R}}}}^{z_{l}}(t)=\! -\frac{{i}}{\hbar}\!\sum_{r}\!\boldsymbol{{A}}^{z_{l}\dagger}(t)\boldsymbol{{h}}_{r}^{z_{l}}(t)\boldsymbol{{A}}^{z_{l}}(t)\boldsymbol{{R}}^{z_{l}}(t)\orthrep{{\boldsymbol{{H}}}}_{r}^{z_{l}}(t).
\label{eq:psi_eom_2}
\end{equation}
linearized forms of these equations form the basis of the
PSI approach. When combined with the serial updating scheme that will
be discussed in the companion paper, this approach provides a second
order integration scheme for evolving the ML-MCTDH EOMs, in which
the EOMs that are solved are free of singularities. These two linearized
EOMs correspond to the time-dependent Schr\"{o}dinger equation expanded
in terms of a time-independent, incomplete and orthonormal basis. Whenever
$\boldsymbol{{R}}^{z_{l}}$ is singular, some of these basis functions
become arbitrary. While this does not effect the accuracy of the expanded
wavefunction when the transformation is performed, it could lead to
sub-optimal performance of approaches based on these EOMs. This issue
arises from exactly the same source as the singularities in the standard
ML-MCTDH EOMs. When the wavefunction is rank deficient the ML-MCTDH wavefunction
contains terms that do not contribute to the overall value of the
fully expanded wavefunction. These terms are arbitrary and linear
variational principles do not define the evolution of such terms.

In contrast to the singularities present in the ML-MCTDH
approach, which lead to significant numerical instabilities that persist
past the point where the wavefunction is rank deficient and where the mean-field
density matrix is exactly singular, the construction of basis functions
satisfying all of the conditions imposed by the linear variational
principle is a stable albeit arbitrary process. Once the wavefunction
is not rank deficient, this is no longer the case and the construction
of these basis functions is well-defined. All of EOMs derived in this manner are non-singular
provided the constraints applied are free of singularities, and
as a result the equations do not suffer from issues due to ill-conditioning.
At no point is it necessary to employ regularization and these approaches
therefore entirely remove the need to specify a regularization parameter,
a parameter that is not related to the physics of the problem and that can 
significantly effect the accuracy and efficiency of simulations.

When the ML-MCTDH wavefunction is nearly rank deficient, the transformed coefficient 
tensors, and hence the transformed SPFs, do not exhibit the same rapid evolution that is 
observed for the standard coefficient tensors in standard ML-MCTDH.   Upon transforming to the standard coefficient tensors, however, 
we observe that the evolution of the transformed SPFs corresponds to a 
rapid evolution of the standard SPFs.  This rapid
evolution arises due to the choice of a nearly linearly dependent (incomplete) set of basis functions.  When represented via an orthonormal set of functions 
spanning the same space, this evolution is not rapid.  As such, 
it should be possible to use considerably larger time steps in the evolution of these transformed SPFs. 
When the wavefunction is rank deficient, these unoccupied SPFs become arbitrary.
As a consequence, the dynamics may depend on the selection made for these 
arbitrary functions, namely on terms that do not contribute to evaluation of
any observable properties. In practice, we find that this does not often
appear to be a significant of an issue and becomes less significant
as the number of SPFs at each node is increased. We will consider
this important point in more detail in the companion paper.

\section{The Invariant Equations of Motion\label{sec:The-Invariant-Equations}}
Recently, a new approach for time-evolving ML-MCTDH ansatz
wavefunctions has been presented in Ref. \onlinecite{Weike2021}. This approach
shares many features with the standard ML-MCTDH approach and with
the PSI approach. It employs a parallel constant mean-field (CMF)
scheme, as is often done for the standard ML-MCTDH EOMs. Like the EOMs
presented above, it makes use of the representation of the wavefunction
given in Eq.\,\ref{eq:transformed_wavefunction_representation}, and
involves propagating EOMs for transformed coefficient tensors. However,
one key difference between the invariant EOMs and the singularity
free EOMs is evident in how they treat the additional dynamic gauge freedom
introduced by the decomposition in Eq.\,\ref{eq:SHF_decomposition}.
Rather than applying a dynamic constraint on the transformed SHFs,
the invariant EOMs remove the gauge freedom by requiring $\boldsymbol{{R}}^{z_{l}}(t)=\boldsymbol{{\rho}}^{z_{l}}(t)^{\frac{{1}}{2}}$ for all times,
where the square root is taken to be the principle square root.\citep{Weike2021}
Via this requirement, the EOMs for $\boldsymbol{R}^{z_{l}}(t)$
have been fully specified, and as such this constraint fixes both the
static and dynamic gauge for the transformed SHFs. While the choice
to use the principle square root does lead to appealing properties for
the $\boldsymbol{{R}}^{z_{l}}(t)$, e.g. it is Hermitian for all time,
it is entirely arbitrary. 

Making this choice of constraint
for $\boldsymbol{R}^{z_{l}}(t)$ is equivalent to imposing the following
dynamic gauge condition on the transformed SHFs
\begin{equation}
\dotorthrep{{\boldsymbol{{\Psi}}}}{}^{z_{l}}(t)\orthrep{\boldsymbol{\Psi}}^{z_{l}\dagger}(t)=-\frac{{i}}{2}\boldsymbol{{g}}^{z_{l}}(\boldsymbol{{R}^{z_{l}}}(t)).\label{eq:invariant_gauge_condition}
\end{equation}
This can be shown by expanding the EOMs for the standard
SHFs in terms of this new representation, and inserting the EOM for
the $\boldsymbol{R}^{z_{l}}(t)$. The constraint matrix, $\boldsymbol{{g}}^{z_{l}}(\boldsymbol{{R}}^{z_{l}}(t))$,
is a complicated function of $\boldsymbol{R}^{z_{l}}(t)$ that requires
the evaluation of inverse of the superoperator, $\boldsymbol{{\mathcal{{R}}}}^{z_{l}+}$
defined by
\begin{equation}
\boldsymbol{{\mathcal{{R}}}}^{z_{l}+}\boldsymbol{{O}}=\boldsymbol{{R}}^{z_{l}}\boldsymbol{{O}}+\boldsymbol{{O}}\boldsymbol{{R}}^{z_{l}},
\end{equation}
(see Eq. 62 of Ref. \onlinecite{Weike2021}). Whenever $\boldsymbol{R}^{z_{l}}(t)$
is singular, that is whenever the ML-MCTDH wavefunction is rank deficient,
this constraint matrix is unbounded. 

In arriving at the invariant EOMs approach it
is necessary to also chose a gauge condition for the SPFs. In Ref.
 \onlinecite{Weike2021}, this condition is selected so that the transformed
SPFs and standard SHFs evolve according to the same EOMs. 
This condition is satisfied for all members of the singularity
free EOMs where $\boldsymbol{x}^{z_{l}}(t)=\boldsymbol{y}^{z_{l}}(t)$.
For the invariant EOMs, this therefore requires
setting the dynamic constraint matrix for the SPFs to
\begin{equation}
\boldsymbol{{\Phi}}^{z_{l}\dagger}(t)\dot{{\boldsymbol{{\Phi}}}}{}^{z_{l}}(t)=-\frac{{i}}{2}\boldsymbol{{g}}^{z_{l}}(\boldsymbol{{R}}^{z_{l}}(t)).
\end{equation}
With these conditions, the transformed coefficient tensors evolve
according to the invariant EOMs
\begin{equation}
\dotorthrep{\boldsymbol{A}}^{z_{l}}=-\frac{{i}}{\hbar}\sum_{r}\boldsymbol{h}_{r}^{z_{l}}\orthrep{{\boldsymbol{{A}}}}{}^{z_{l}}\orthrep{{\boldsymbol{{H}}}}_{r}^{z_{l}}+\frac{{i}}{2}\orthrep{{\boldsymbol{{A}}}}{}^{z_{l}}\boldsymbol{{g}}^{z_{l}}+\frac{{i}}{2}\left(\bigoplus_{k}\boldsymbol{{g}}^{z_{l},k}\right)\orthrep{{\boldsymbol{{A}}}}{}^{z_{l}}.\label{eq:invariant_eoms}
\end{equation}
This equation has exactly the same form as Eq.\,\ref{eq:sf_eom_1},
but with a specific choice for the SPF and transformed SHF constraint
matrices, namely they are both equal to $\boldsymbol{{g}}^{z_{l}}(\boldsymbol{{R}}^{z_{l}}(t))$.
Similarly, the $\boldsymbol{{R}}^{z_{l}}(t)$ matrices evolve according
to EOMs with the same form as Eq.\,\ref{eq:sf_eom_2}. 

Having shown that the invariant EOMs constitute a particular choice of gauge-constraint in the framework of singularity-free EOMs introduced above, we will briefly discuss some consequences of this choice. First, although the right hand side of Eq.\,\ref{eq:invariant_eoms}
is always bounded,\citep{Weike2021} it can potentially involve the
evaluation of an unbounded operator acting on the transformed coefficient
tensors. When it is linearized, as is implicitly done in Ref. 
\onlinecite{Weike2021}, the solution of the linearized equation requires
the evaluation of the exponential of a matrix that is unbounded whenever
$\boldsymbol{{R}}^{z_{l}}(t)$ is singular. It is therefore necessary
to regularize these equations, bounding this operator and allowing
for the evaluation of regularized dynamics. The fact that the right
hand side of Eq.\,\ref{eq:invariant_eoms} is bounded thus does not alter this fact.

Second, the derivation of the invariant EOMs relies on the orthonormality of the transformed SHFs. However this cannot be guaranteed when using the constraint $\boldsymbol{{R}}^{z_{l}}(t)=\boldsymbol{{\rho}}^{z_{l}}(t)^{\frac{{1}}{2}}$. 
Whenever any of the singular values of $\boldsymbol{{\rho}}^{z_{l}}(t)$ are small it becomes necessary to use a regularized inverse in the construct of the transformed SHFs.\citep{Weike2021}  This results in non-orthogonal transformed SHFs and in turn 
the overlap matrix of the transformed SHFs, which can be interpreted as a transformed
mean-field density matrix, is not the identity matrix. In these situations, Eq.\,\ref{eq:invariant_eoms} is an approximation to the correct EOMs.  Further, whenever $\boldsymbol{{\rho}}^{z_{l}}(t)$ is exactly singular the transformed mean-field density matrix is also singular, and so the EOMs including this term will suffer from the same 
 issues involving the inverse of singular matrices as is the case for the standard ML-MCTDH EOMs. 
The invariant EOMs are only free of the singularities arising from the inverse mean-field density matrix if this matrix is ignored.

Having said this, the invariant EOMs have been demonstrated
to perform better than standard ML-MCTDH for a set of spin-boson models.\citep{Weike2021}
For these models, it was demonstrated that a small regularization
constant is not necessary to accurately capture the evolution of the
initially unoccupied SPFs. As we will show in the companion paper,
this can be taken further, and no regularization is required at all.
For the problems considered in Ref.\,\onlinecite{Weike2021}, the potential
issues we have raised here were not observed. Whether these
issues become significant for other parameter regimes and in different models is unclear.

The approach taken to solve the equations in Ref.
\onlinecite{Weike2021} differs significantly from the PSI.
In particular, this approach is completely parallelizable and the transformed
coefficient tensors for all nodes can be updated in parallel using
a constant mean-field (CMF) integration scheme.  As we have discussed
above, there is nothing preventing such an integration scheme being
applied to any of the singularity-free EOM approaches. As such, the PSI EOMs can
likewise be solved using an integration scheme of this form, and we present
a brief discussion and numerical results for such an approach in Sec. II of the
Supplementary Information.
Recently, there have been developments in employing parallel algorithms for 
integrating the PSI EOMs for MCTDH wavefunctions  \cite{Ceruti2021a}.  The approach of Ref.~\onlinecite{Ceruti2021a}
and the scheme presented in Ref. \onlinecite{Weike2021} share a number of
similarities, in particular they both use similar strategies for avoiding the backward 
in time evolution of the $\boldsymbol{R}^{z_l}$ matrices.
 Apart from the different choices gauge constraints, the main difference between the two algorithms is the use of a rescaled projection in Ref.~\onlinecite{Weike2021}, while no rescaling is applied in Ref.~\onlinecite{Ceruti2021a}. 
A discussion of practical parallel integration schemes and their numerical
performance is left to future work.  

\section{Conclusion\label{sec:Conclusion}}
In this paper, we have rederived the EOMs that form
the basis of the PSI using a framework
that is closely related to that used for deriving the standard ML-MCTDH
EOMs.  In doing so we highlight the fact that the key differences between the
projector splitting integrator EOMs, the standard ML-MCTDH EOMs, and
the invariant EOMs originate from the choice of gauge and the consequent constraints
placed on the evolution of the coefficient tensor. For the standard
ML-MCTDH approach, the gauge choice constrains the coefficient tensor
to be semi-unitary at all times. As a consequence, non-unitary
factors in the ML-MCTDH expansion of the wavefunction are absorbed
into the SHFs and the ML-MCTDH EOMs become singular whenever the
SHFs are linearly dependent. For both the PSI EOMs and invariant EOMs,
an alternative representation is used in which the non-unitary factors
are moved from the SHFs to the coefficient tensors. These two approaches
differ in their choice of dynamic gauge conditions. The invariant
EOMs are obtained when the non-unitary factors in the wavefunction
are constrained such that they are Hermitian and are
given by the principle square root of the mean-field density operator
at all times, while a specific dynamic constraint is placed on the SPFs.
These choices introduce a potentially singular contribution to the evolution
of the coefficient tensor and require the use of a regularized
inverse in the construction of the transformed SHFs. When regularization
is required, this does not correctly transfer the non-unitary factors
away from the SHFs and thus may lead to issues with the accuracy of
the approach. In contrast, the PSI employs constraints on
the dynamics of the SPF and transformed SHFs with no constraints placed
on the non-unitary factors that are evolved. The resultant
EOMs are always well defined and numerically well-behaved. They simply
correspond to the Schr\"{o}dinger equation expanded in terms of the orthonormal
but incomplete basis defined by the SPFs and transformed SHFs for
a node. We have further shown that the PSI EOMs and invariant EOMs
are simply two specific cases of a wide family of EOMs. Provided the
gauge conditions selected are non-singular, as is the case for the
PSI, these EOMs are non-singular. Any method based on the solution
of these equations does not require the use of regularization. As such,
approaches based on these equations do not require the specification
of a parameter that can considerably alter the
efficiency and accuracy of the dynamics. 

While all numerically exact methods that make use of the
ML-MCTDH ansatz for the wavefunction and use the time-dependent variational
principle to perform time-propagation can become ill-defined whenever
the wavefunction is rank deficient, rank-deficiency manifests in different ways
for different approaches. For the standard ML-MCTDH approach it
manifests in the singularities in the EOMs discussed above. For the
PSI approach it manifests as non-uniqueness in the choice of the
incomplete basis functions. For the invariant EOMs the consequences of rank-deficiency
manifest as both type of difficulties. While these two sets of issues have the same
origin, they lead to considerably different behavior of the approaches in practice.
In particular, the use of regularization to handle the issues associated with singularities
in the EOMs introduces issues that can persist past the point in which
the wavefunction is rank deficient. It is then necessary to converge the
dynamics with respect to the regularization parameter, and there is
no guarantee that this convergence can be obtained. The issues associated
with the non-uniqueness in the choice of incomplete basis functions
appear to be less severe. The construction of a set of basis functions that
satisfy the conditions required is numerically stable. In the limit of a full basis
set expansion this non-uniqueness is not relevant as the basis functions
necessarily span the full space, and as such the issue of completeness can be viewed as a
convergence issue. In the companion paper, we will discuss this in
more detail. We will discuss the numerical implementation of the multi-layer
form of the PSI algorithm, with an aim of connecting the steps involved to
equivalent steps needed for the implementation of standard ML-MCTDH. 
From this vantage point we will present a number of applications of the PSI approach
to a series of challenging model problems which illustrate its advantages over ML-MCTDH
schemes which require regularization.
\begin{acknowledgments}
L.P.L. and D.R.R. were supported by the Chemical Sciences, Geosciences and Biosciences Division of the Office of Basic Energy Sciences, Office of Science, U.S. Department of Energy.
\end{acknowledgments}

\section*{Data Availability}
The data that support the findings of this study are available from the corresponding author upon reasonable request.

\bibliography{part_1}

\end{document}